\documentclass[aps,pre, reprint,superscriptaddress, twocolumn,amsmath,amssymb,nofootinbib,showpacs]{revtex4}

\usepackage{amsmath}
\usepackage{graphicx}
\usepackage{amssymb}
\usepackage{dcolumn}
\usepackage{bm}
\usepackage{color}

\makeatletter

\begin{document}

\title{Escape rate of an active Brownian particle over a potential barrier}

 \author{P.S. Burada} \thanks{Corresponding author: {\tt burada@pks.mpg.de}}
 \affiliation{Max Planck Institute for the Physics of Complex Systems,  N\"othnitzer Str. 38, 01187 Dresden, Germany}
 \affiliation{Georg-August-Universit\"at G\"ottingen, Institut f\"ur
   Theoretische Physik, Friedrich-Hund-Platz 1, 37077 G\"ottingen, Germany}

\author{B. Lindner$^{1,}$}\thanks{\tt benjamin.lindner@physik.hu-berlin.de}
\affiliation{Bernstein Center for Computational Neuroscience Berlin \& 
Physics Department, Humboldt University Berlin, Philippstr. 13, Haus 2, 10099 Berlin, Germany}

\begin{abstract}

\centerline{\today}

We  study the dynamics of an active Brownian particle with a nonlinear friction function located in a spatial cubic potential.  
 For strong but finite damping, the escape rate of  the particle  over the spatial potential barrier shows a nonmonotonic dependence on the noise intensity.  We relate this  behavior  to the fact that the active particle escapes from a limit cycle rather than from a fixed point and that a certain amount of noise can stabilize the sojourn of the particle on this limit cycle.   
\end{abstract}


 \maketitle


The escape of a Brownian particle out of a metastable potential well, the famous Kramers problem, 
is  a long studied subject in statistical physics \cite{Kramers}
with applications   in physics, physical chemistry, biology, and other fields. 
This is so because Brownian motion in a nonlinear force field  can be also employed as a model for the reaction coordinate of a chemical reaction 
(the original problem Kramers was interested in), the phase difference in a  Josephson junction \cite{Ris84},   
or the membrane potential of a nerve cell \cite{GerKis02}, to name but a few prominent  examples. 

Over the last decade another class of stochastic models, so-called active Brownian motion, has attracted much attention 
\cite{SchGru93,SchEbe98,ErdEbe00,Sch02, Lin08}.  
In order to describe self-propelled or active motion of particles in 
biology \cite{Bray,Parrish}, such as the motility of objects at the subcellular level (e.g., assemblies of molecular motors) 
\cite{Howard,molmot}, of cells \cite{SelLi08}, or even of
entire flocks of animals \cite{flocks}, theoreticians  
have used Langevin equations that are endowed with a speed-dependent friction coefficient.
Essential for an active particle is not the mere speed dependence of the friction 
coefficient that can be also observed in equilibrium models
\cite{Lin07}, but also that the friction coefficient is negative over a  range of velocities.

Active Brownian motion has been predominantly studied for the cases of free motion of single  and coupled particles and  for active particles subject to harmonic potential forces.  How an active Brownian particle escapes from a metastable potential 
well has received comparably little attention \cite{EbeLSG05}. 
In this Brief Report, we study this interesting  generalization of the Kramers problem to active particles in the case of  strong but finite damping.  We show that the escape rate can display a   nonmonotonic dependence on the noise intensity, a behavior that is in marked contrast to the classical Kramers rate, which always increases with growing noise level. 


{\it Model.\,\,--} 
The dynamics of an active Brownian particle with a nonlinear friction function $\gamma(v)$
and subject to a spatial potential $U(x)$ reads
\begin{eqnarray}
  \label{eq:Langevin}
  \dot{x}  = v, \,\,\, m \dot{v}   = -\gamma(v) v - \gamma_0 U^{\prime}(x) + \sqrt{2 D} \, \xi(t) \,,
\end{eqnarray}
where $m$ is the mass of the particle, 
$\gamma(v)$ is the nonlinear friction function (see below), and
$\xi(t)$ is Gaussian white noise with  noise strength $D$ 
and  correlation function $\langle \xi(t)\,\xi(t') \rangle = \delta(t - t')$.
The spatial potential is $U(x) = A\left(x - \frac{x^3}{3}\right)$ with the amplitude $A$.
The potential possesses a minimum at $x=-1$ and a maximum (the
barrier) at $x=1$.   
Equation~(\ref{eq:Langevin}) is integrated with a simple  stochastic Euler scheme for which we use a time step of $\Delta t = 0.001$.

Note that we scale the potential force in Eq.~(\ref{eq:Langevin}) by
a parameter $ \gamma_0$, which we will also use as the amplitude of
the nonlinear friction function  below.  In the non-equilibrium
problem considered here, the strength of the active friction function
is not related  with the Stokes friction coefficient, but arises as a
phenomenological parameter, which in our scaling of the spatial potential force conveniently quantifies the time-scale ratio of $x$ and $v$. In the following, we will consider two different nonlinear friction functions, namely, the Rayleigh-Helmholtz (RH) friction function  and the Schweitzer-Ebeling-Tilch (SET) friction function \cite{SchEbe98}, and study the  escape dynamics of the active particle over the potential barrier.

The RH friction function reads 
$\gamma(v) = \gamma_0\left(v^2 - u_0^2\right)$.
Here, $u_0>0$ is the speed the particle would attain in the long-time limit if noise and potential would be switched off. 
In contrast to passive Brownian motion, a vanishing velocity is
dynamically unstable because the friction function is negative for $-u_0<v<u_0$.  
The SET friction function is given by 
$\gamma(v) = \gamma_0\left(1 - \frac{\beta}{1 + v^2}\right)$.
With this friction function, Brownian particles attain a self-propelled motion if $\beta > 1$.
 
For the considered models, the nullcline for the fast variable $v$ is
the line (or the lines)  at which $\dot{v}=0$ in the absence of noise,
determined by $\gamma(v)\, v + \gamma_0\,A (1-x^2) = 0 \,$.
If we regard this as a cubic equation in $v$,  we see that for a
given $x$, one, two, or three solutions for the velocity are possible.
If there are three solutions, the middle one is dynamically unstable
 (in $v$), while the upper and the lower branches attract
 trajectories. The intersections of the $v$ nullcline with  the $x$ nullcline
 ($v=0$) form only two unstable fixed points.
The critical amplitude $A_c$  at which the bifurcation occurs 
can be calculated. For the RH model it is 
$A_c = 2 \,u_0^3/(3^{3/2}) \;\;\; (\approx0.38\,\,\text{for}\,\, u_0 = 1)$ 
and for the SET model it reads
$A_c =[(3\beta-d)/(d-\beta)] \sqrt{(d-2-\beta)/2} \;\;\;
(\approx 0.3\,\, \text{for}\,\, \beta = 2)$, 
where $d=\sqrt{\beta(8 + \beta)}$. 
For the simulations, if not stated otherwise, we set the parameters $m=1$, $u_0=1$, $\beta=2$, and $\gamma_0=20$.

\begin{figure}[t]
  \includegraphics[scale= 0.55]{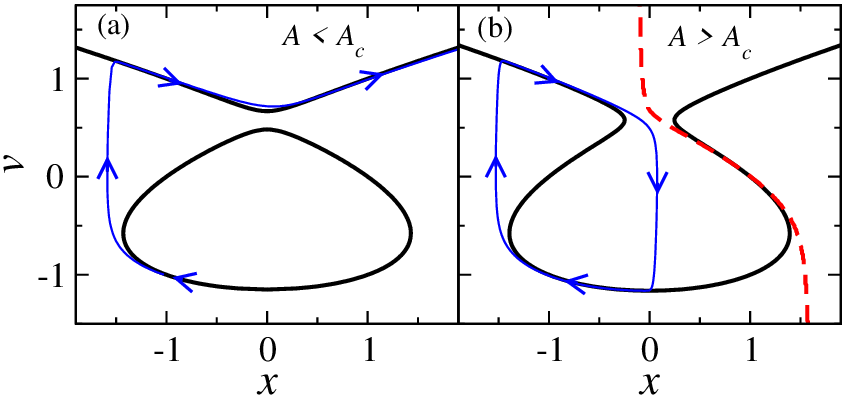} 
  \caption{(Color online) Deterministic behavior below (a) and above
    (b) the critical amplitude for the RH model. 
    The red dashed line in (b) is the separatrix line calculated from the deterministic equation.
   Qualitative similar behavior is observed for the SET model.}
  \label{fig:deterministic}
\end{figure}
Typical trajectories of the deterministic system are shown in Fig.~\ref{fig:deterministic} for potential amplitudes that are below [Fig.~\ref{fig:deterministic}(a)] and above [Fig.~\ref{fig:deterministic}(b)] the critical value $A_c$. For a small amplitude ($A<A_c$), the trajectory follows closely the $v$ nullcline and escapes quickly from the potential minimum.
For a large amplitude, in contrast, the particle remains in the minimum and approaches a limit cycle, similar to what was 
observed for active Brownian motion in a harmonic potential \cite{ErdEbe00}.
Because we have chosen $\gamma_0$ to be large, the particle performs relaxation oscillations and switches rapidly between the two stable branches of the $v$ nullcline. 
In Fig.~\ref{fig:deterministic}(b) we have  marked the separatrix (red dashed line), which limits the region of attraction for the limit cycle (the line was determined 
numerically from deterministic simulations with different initial conditions).  Put differently, deterministic trajectories started to the right (left) of  the separatrix line will ultimately escape from the minimum (end up on the limit cycle). 

\begin{figure}[t]
  \includegraphics[scale= 0.6]{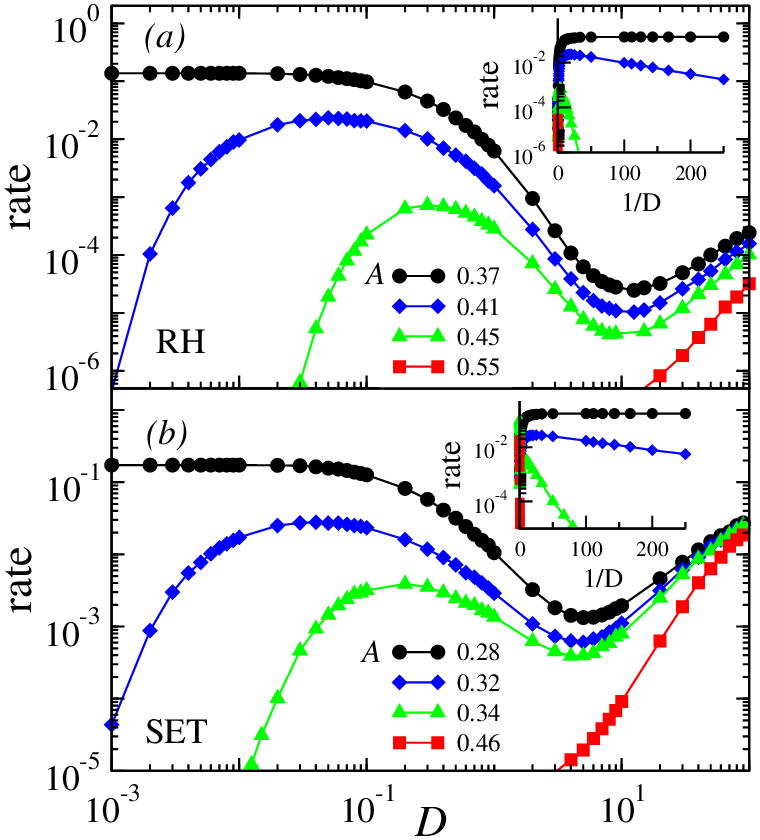} 
  \caption{(Color online) 
    Escape rate as a function of the noise strength for  
    (a) RH and (b) SET models at different amplitudes of the spatial potential. 
    For the RH model the critical $A$ value is $ 0.38$ and for the SET model it is $0.3$.
    Insets depict Arrhenius plots of the same data.}
  \label{fig:kramers}
\end{figure}

{\it Escape statistics.\,\,--}
In order to measure the escape rate, we keep the particle in the potential minimum $x(t=0)=x_0=-1$ 
with a negative velocity, $v(t=0)=v_0=-1$, a position in phase space that is always on the $v$ nullcline. 
We measure the time it takes the particle to overcome a threshold of $x_{th}=5$ and repeat this numerical 
experiment $2000$ times. The rate is then given by the inverse of the mean first passage time from 
$(x_0,v_0)=(-1,-1)$ to the threshold in $x_{th}$ (regardless of
the velocity with which the particle arrives at $x_{th}$).  
Other initial conditions, as long as they are chosen left of the
separatrix, do not strongly affect our simulation results.

The escape rate of active Brownian particles is depicted in
Fig.~\ref{fig:kramers} for both models. 
A peculiar,  non-monotonic behavior of the escape rate can be observed. We recall that for passive Brownian particles, in a spatial potential, the escape rate always increases monotonically 
with the noise strength $D$ \cite{Kramers}. For the active particle, remarkably,
 for a potential amplitude $A$ sufficiently exceeding the critical one and over a range of $D$ the escape rate decreases with increasing $D$, giving rise to a maximum at intermediate  and a minimum at large noise intensity, respectively.  

For large amplitude, the escape rate shows at very small noise an Arrhenius-like behavior $r\sim \exp(-\Delta/D)$  as can be seen in  the logarithmic plots vs $1/D$ in the insets of  Fig.~\ref{fig:kramers}. In the opposite limit, for a small potential amplitude $A<A_c$, the particle is driven out of the potential minimum by the deterministic dynamics as was already discussed in the context of Fig.~\ref{fig:deterministic}(a). Consequently, the 'escape rate' in this case is finite and decreases for increasing noise; this is also the only case considered here, in which the rate depends on the specific initial conditions. 

The maximum in the rate is most pronounced for $A=0.41$ [i.e., a potential amplitude close to but exceeding the critical amplitude  (and also its modified value for finite $\gamma_0$)]. Larger amplitudes generally reduce the rate for all noise intensities but particularly at weak noise. Furthermore, the location of the maximum shifts to larger values of $D$. In the limit of very large potential amplitude $A$, the maximum of the rate vanishes and the rate strictly increases with increasing noise intensity.  All of this applies to both friction models, which indicates that the rate maximization at finite noise is a robust phenomenon  that does not hinge on the fine details of the model.  
 
\begin{figure}[t]
  \includegraphics[scale=0.6]{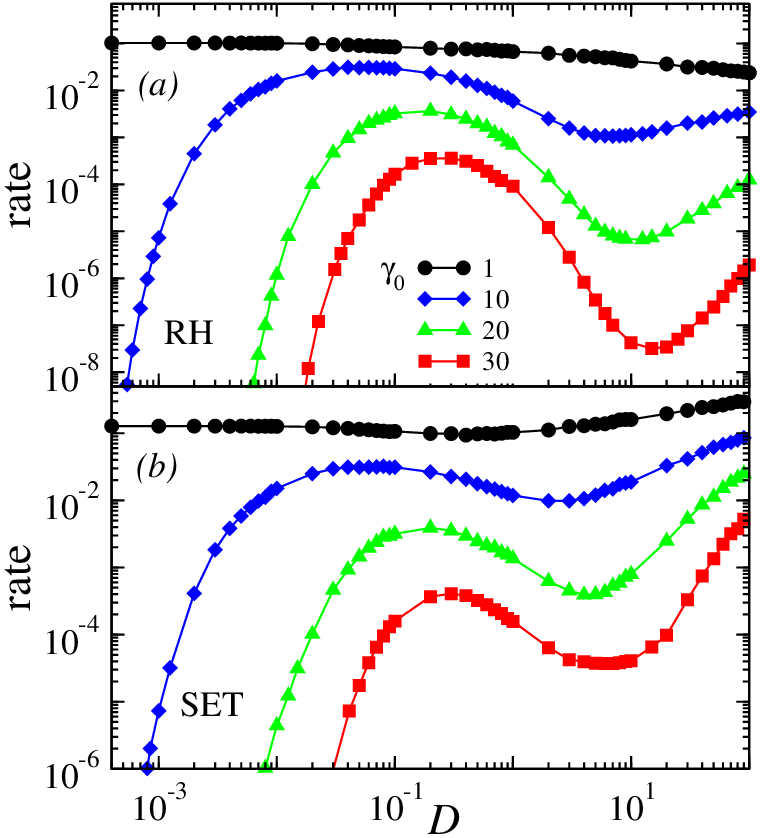} 
  \caption{(Color online) 
    Escape rate as a function of the noise strength for  
    (a) RH and (b) SET models at different strengths $\gamma_0$ of the nonlinear friction function. 
    Here, we set the amplitude of the spatial potential $A = 0.43$ for the RH model and 
    $A = 0.34$ for the SET model.}
  \label{fig:kramers_g}
\end{figure}

Figure~\ref{fig:kramers_g} shows the behavior of the escape rate 
for different amplitude  $\gamma_0$. We recall that $\gamma_0$ in our scaling of the potential determines the time scale between $x$ and $v$.
For larger $\gamma_0$, the intrawell limit-cycle oscillations turn into pronounced relaxation oscillations. At the same time, the escape over the barrier also becomes more difficult because the particle is strongly attracted to the stable branches of the $v$ null-cline.  As a consequence, we observe for increasing values of $\gamma_0$ 
an overall reduction in the rate and shift of the local rate maximum toward larger noise intensity. For large values of $\gamma_0$ and for very weak noise, we observe again a Arrhenius-like behavior as was already illustrated by the insets in Fig.~\ref{fig:kramers}. Going to the opposite limit of moderate-to-small $\gamma_0$ (e.g., $\gamma_0=1$), the maximum vs noise intensity vanishes and decay of the rate in the weak-noise limit turns into a saturation. Hence, a pronounced but not perfect time-scale separation between $x$ and $v$ seems to be precondition for a  maximum of the escape rate.

\begin{figure}[t]
 \includegraphics[scale=0.6]{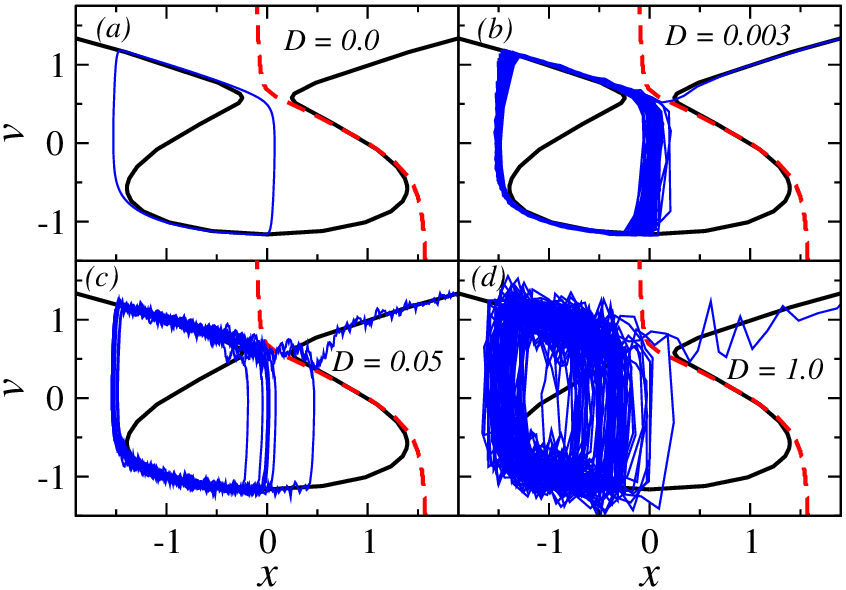}
 \caption{(Color online) 
   Escape trajectories at different noise strengths
   for $\gamma_0 = 20$ and $A=0.41$.
   The black lines are nullclines and the red dashed line is  the separatrix line
   of the deterministic motion in Eq.~\ref{eq:Langevin}.}
 \label{fig:escape}
\end{figure}

In order to understand the peculiar dependence of the escape rate on the noise
strength  for $A>A_c$, we now analyze the escape statistics in more
detail,  restricting ourselves to the RH model. 
The dynamics is illustrated by sample trajectories for various noise levels in Fig.~\ref{fig:escape}. 
In the deterministic case [$D=0$, Fig.~\ref{fig:escape}(a)], the
particle sticks to the limit cycle and cannot escape over the barrier.  With a finite but small amount of noise [Fig.~\ref{fig:escape}(b), e.g., $D=0.003$], the particle follows the limit cycle for 
many rounds but  can finally escape across the separatrix through the bottleneck around $x=0$  and $v=1/2$.
A further increase of $D$  [e.g., to the level $D=0.05$ shown in Fig.~\ref{fig:escape}(c)] increases the probability to escape already after only a few rounds on the limit cycle. 
At this specific value the escape rate is maximum and its growth with increasing noise is so far not surprising.  However, if we further increase the noise, we stabilize the motion on the limit cycle and thus reduce the probability of escape 
through the bottleneck by the following mechanism. As the particle passes along the upper stable branch of the $v$ nullcline,  a competing escape event becomes feasible, namely, the early jump to the lower branch of the $v$ nullcline.  Strong noise increases this switching rate  and, consequently, keeps the particle from reaching the critical region close to the separatrix. 

\begin{figure}
  \includegraphics[scale=0.6]{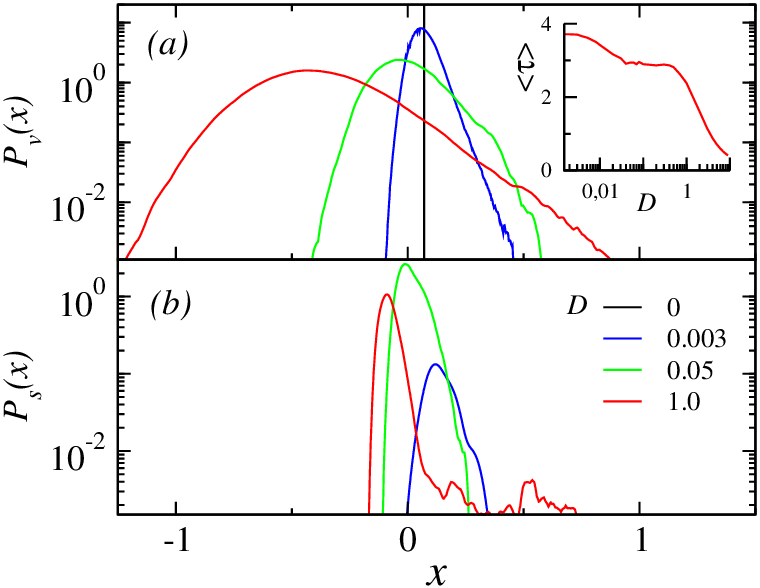} 
  \caption{ (Color online) 
  (a) probability distribution of points of switching between the upper and lower stable branches of the $v$ null-cline; (b) probability distribution of escape points across the separatrix  region.  Inset in (a): mean cycling time  vs  noise intensity. Parameters are same as in Fig.~\ref{fig:escape}. }
  \label{fig:prob}
\end{figure}

Further support for this mechanism comes from the statistics of the switching points $x$ at which the particle either  (i) crosses the line $v=0$ when jumping from the upper stable branch of the $v$ nullcline to the lower one [probability $P_v(x)$ in Fig.~\ref{fig:prob}(a)] or  (ii) crosses the separatrix and escapes from the potential minimum [probability $P_s(x)$ in Fig.~\ref{fig:prob}(b)].   We normalize the histograms for both densities in $x$ to the total
number of events (i.e., $\int_{-\infty}^{x_{th}} dx [P_v(x)+P_s(x)]=1$).
In addition,  we also calculate $\langle \tau \rangle$, which is  the mean time difference between two successive turns on  the limit cycle [mean cycling time, shown in  the inset in Fig.~\ref{fig:prob}(a)].
As demonstrated in  Fig.~\ref{fig:prob}, the switching distribution $P_v(x)$  broadens considerably when we reach the range of noise values where the escape rate drops with increasing noise, indicating a substantial increase in the probability of early transitions from upper to lower branch. In the same range of noise values the mean cycling time   $\langle \tau \rangle$ decreases only little  and the distribution of escape points over the separatrix remains for all noise levels located  in a narrow region around $x=0$.   We take this as an indication that the stabilizing effect of noise is mainly related to the switching on the limit cycle and not so much to  the escape once the particle has reached the vicinity of the separatrix. 

{\it Summary.\,\,--} 
We have shown that the escape dynamics of active Brownian particles, with a nonlinear 
friction function, in a spatial cubic potential is distinctly different from the passive case. 
We have found that the combination of two nonlinearities, in space and velocity, gives rise to a non-monotonic  escape rate as a  function of the noise strength, an effect that robustly occurs for different 
friction functions. The maximum in the rate could be understood by the stabilizing effect of noise on the dwell time in the potential minimum. 

As the model studied here has been shown to qualitatively approximate the dynamics of coupled molecular motor systems \cite{molmot}, it would be an interesting task to study the escape problem for the latter system. 
Furthermore, the non monotonic rate dependence on noise is most likely
not restricted to one-dimensional models but could be also expected in
higher dimensions. Particularly interesting should be the dependence
in systems in which the existence of metastable states hinges on the
presence of nonlinear friction (see, e.g.,  the Toda chain studied in
Ref. \cite{Dunkel}).


This research was supported by the Max Planck Society and the German
Federal Ministry of Education and Research (BMBF), Grant No. FKZ:01GQ1001A.

\end{document}